\documentclass{article}  
\usepackage{amssymb}
\usepackage{graphicx}
\usepackage{xy} \xyoption{all}

\newenvironment{pagi}[1]{\begin{minipage}[t]{#1em}}{\end{minipage}}

\newtheorem{prop}{Proposition}[section]

\newcommand{\C}{\mathbb{C}}

\newcommand{\hache}{\mathbb{H}}
\newcommand{\I}{\mathbb{I}}

\newcommand{\N}{\mathbb{N}}

\newcommand{\R}{\mathbb{R}}
\newcommand{\U}{\mathbb{U}}
\newcommand{\V}{\mathbb{V}}

\newcommand{\Z}{\mathbb{Z}}

\newcommand{\pint}[2]{\left\langle #1|#2 \right\rangle}

\newcommand{\tr}{\mbox{\rm Tr}}

\newcommand{\vcg}[1]{\mbox{$\mathbf{#1}$}}
\newcommand{\vcgs}[1]{\mbox{\scriptsize $\mathbf{#1}$}}
\newcommand{\id}[1]{\I_{#1}}
\newcommand{\imb}{i}
\newcommand{\drc}[1]{\left|#1\right\rangle}
\newcommand{\tri}[2]{\mbox{\rm Tr}_{#1}\left( #2\right)}

\title{Embeddings of spaces of quregisters into special linear groups}

\author{Dalia Cervantes and Guillermo Morales-Luna \\
Computer Science Department, CINVESTAV-IPN, Mexico City, Mexico \\
dalia@computacion.cs.cinvestav.mx, gmorales@cs.cinvestav.mx}

\date{\today}
\begin{document}
\maketitle

\begin{abstract}
We study embeddings of the unit sphere of complex Hilbert spaces of dimension a power $2^n$ into 
the corresponding groups of non-singular linear transformations. For the case of $n=1$, the sphere $S_1(\C)$ of qubits
is identified with $\mbox{\rm SU}(2)$ and the algebraic structure of this last group is carried into $S_1(\C)$. 
Hence it is natural to analyse whether is it possible, for $n\geq 2$, to carry the structure of the symmetry group $\mbox{\rm SU}(2^n)$ into the unit sphere $S_{2^n -1}(\C)$.
For 
$n=2$ the embeddings of  $S_3(\C)$ into $\mbox{\rm GL}(2^2)$, obtained as tensor products of the above embedding, fails to determine a bijection between $S_3(\C)$ and $\mbox{\rm SU}(2^2)$, but they determine entanglement measures consistent with von Neumann entropy.
\end{abstract}

 \section{Introduction}
The basic information particle in Quantum Computing is the so called {\em qubit} which can be realised as a normalised linear combination of an ``up'' ($\drc{0}$) and ``down''($\drc{1}$) state for, let us say, the spin of an electron. Formally, the unit 
sphere $S_1(\C)$ of the complex Hilbert space $\hache_1=\C^2$ consists of the {\em qubits}.

The space $\hache_1$ has dimension 2 and its canonical basis is $\{\drc{0},\drc{1}\}$, where
$$\drc{0} = \left[\begin{array}{c} 1 \\ 0 \end{array}\right] = {\bf e}_0\ \ ,\ \ \drc{1} = \left[\begin{array}{c} 0 \\ 1 \end{array}\right] = {\bf e}_1.$$
Any basis $\{{\bf x}_0,{\bf x}_1\}$ of $\hache_1$ is said {\em positively oriented} if the change of basis matrix $[{\bf x}_0\ \ {\bf x}_1]$, with respect to the canonical basis, has determinant equal to 1. Any qubit ${\bf x}_0\in S_1(\C)$ can be associated to a second qubit ${\bf x}_1\in S_1(\C)$ such that $\{{\bf x}_0,{\bf x}_1\}$ is  positively oriented.

The unit circle in the complex plane $\C$ has a natural group structure with complex multiplication. By associating to each qubit 
the positively oriented basis consisting of itself and its orthogonal complement  (in the positive sense), then a natural 
identification of $S_1(\C)$ with $\mbox{\rm SU}(2)$ results. Hence, $S_1(\C)$ inherits the algebraic structure of the symmetries
group $\mbox{\rm SU}(2)$. This suggests the possibility to carry Quantum Computing into the group $\mbox{\rm SU}(2)$. 
However this 
is not possible with the proposed identification because a computer gate $U\in \mbox{\rm U}(2)$ commutes with the embedding $S_1(\C)\to\mbox{\rm SU}(2)$
if and only if $U$ preserves orientation, namely, $\det U = 1$.

The composition of qubits produces more complex information structures. 

Let $\hache_n$ denote the $n$-fold tensor power of $\hache_1$. The elements of its unit sphere 
$S_{2^n -1}(\C)\subset\hache_n$ will be called {\em $n$-quregisters}. The $n$-fold tensor 
product $(S_1(\C))^{\otimes n}$ of $S_1(\C)$ is included in $S_{2^n -1}(\C)$. Hence, the $n$-fold  tensor 
power of qubits are $n$-quregisters. 
Let us call an $n$-quregister {\em separable} if it is the $n$-fold tensor powers of qubits, i. e. 
unit vectors in $\hache_1$.

The main motivation of the current research is to provide the unit sphere of the $2^n$-dimensional complex Hilbert space of an algebraic structure as a homomorphic image of the group structure of the symmetry group $\mbox{\rm SU}(2^n)$.

Naturally, an embedding of the sphere $S_{2^n -1}(\C)$ of $n$-quregisters
into the space of non-singular 
linear transforms $\mbox{\rm SL}(2^n)$ is sought to be congruent with tensor products. This motivates the 
introduction of the maps in relations~(\ref{eq.ps330})-(\ref{eq.ps333}) below, for $n=2$. However, the images of 
the maps meet $\mbox{\rm SL}(2^2)-\mbox{\rm U}(2^2)$, and they are useless to transport the algebraic structure of
either $\mbox{\rm SU}(2^2)$ or $\mbox{\rm U}(2^2)$ into $S_3(\C)$.

Nevertheless, it is possible to introduce a measure of entanglement in $S_3(\C)$ through these maps. 

Several criteria 
have been introduced for entanglement measurement~\cite{Las15,See08,Sab14}. 
The introduced entanglement measure in this paper is consistent with the notion of partial separability through von
Neumann entropy of partial traces.

\section{Preliminaries}\label{Preliminaries}

Let us recall some basic notions.

Let $\U$ be a finite-dimensional complex Hilbert space, and $m=\dim(\U)$, then $\U\approx\C^m$. Let ${\cal L}(\U) = \{T:\U\to\U|\ T\mbox{ is linear}\}$ be the space of linear maps defined on $\U$. Provided with the {\em Hilbert-Schmidt inner product} $(T,S)\mapsto \pint{T}{S} = \tr{T^HS}$, it is a complex Hilbert space of dimension $m^2$. By fixing a basis in $\U$, there is a natural identification ${\cal L}(\U)\approx\C^{m\times m}$: the space of square $(m\times m)$-complex matrices.

An operator $T\in{\cal L}(\U)$ is {\em positive (semidefinite)} if $\exists S\in{\cal L}(\U)$: $T = S^HS$, and it is {\em positive definite} if, besides, it is non-singular. Let $\mbox{\rm Pos}(\U)$ be the collection of positive operators.

A {\em quantum state} is a positive operator $T\in\mbox{\rm Pos}(\U)$ such that $\tr{T}=1$. Let $\mbox{\rm D}(\U)$ be the collection of quantum states.

Now, let $S_{\U} = \{{\bf x}\in\U|\ {\bf x}^H{\bf x}=1\}$ be the unit sphere in $\U$. Clearly, $S_{\U} \approx S_{m-1}(\C)$: the unit sphere in the $m$-dimensional complex linear space. 

The map $\rho:S_{\U}\to\mbox{\rm D}(\U)$, ${\bf x}\mapsto\rho({\bf x}) = {\bf x}\,{\bf x}^H$, is an embedding (namely, an injective map). For each ${\bf x}\in S_{\U}$, $\rho({\bf x})\in{\cal L}(\U)$ is the orthogonal projection along the ray spanned by ${\bf x}$ in the space $\U$. Seen as a matrix,  $\rho({\bf x})$ is called the {\em density matrix} determined by the unit vector ${\bf x}\in S_{\U}$.

Then any unit vector ${\bf x}\in S_{\U}$ can be considered as a state.

A state $T\in\mbox{\rm D}(\U)$ is {\em pure} if $T\in\mbox{\rm Image}(\rho)$, namely, $\exists{\bf x}\in S_{\U}$: $\rho({\bf x})=T$. Non-pure states are called {\em mixed states}, as well.

Let $\V$ be another finite-dimensional complex Hilbert space, with $n=\dim(\V)$. Then the tensor product $\U\otimes\V$ has dimension $mn$.

An unit vector ${\bf u}\in S_{\U\otimes\V}$ is {\em separable} if $\exists ({\bf x},{\bf y})\in S_{\U}\times S_{\V}$: ${\bf u} = {\bf x}\otimes{\bf y}$.

\begin{prop} If $({\bf x},{\bf y})\in\U\times\V$ then $\rho({\bf x}\otimes{\bf y}) = \rho({\bf x})\otimes\rho({\bf y})$.
\end{prop}

An operator $T\in{\cal L}(\U\otimes\V)$ is {\em separable} if there are two sequences $U_0,\ldots,U_{k-1}\in{\cal L}(\U)$,  $V_0,\ldots,V_{k-1}\in{\cal L}(\V)$ such that $T = \sum_{\kappa=0}^{k-1}U_{\kappa}\otimes V_{\kappa}$.

\begin{prop} \label{rm.spsp} For any ${\bf u}\in S_{\U\otimes\V}$, if ${\bf u}$ is separable (as a unit vector) then $\rho({\bf u})$ is separable in $\mbox{\rm D}(\U\otimes\V)$.
\end{prop}
\medskip

Let us consider now the two-dimensional complex Hilbert space $\hache_1=\C^2$ and its
tensor powers, $\forall n>1$: $\hache_n = \hache_{n-1}\otimes\hache_1$. 
Clearly, $\dim \hache_n = 2^n$. The unit sphere $S_{2^n -1}(\C)$ of $\hache_n$ is the set of $n$-quregisters.

Let  $Q=\{0,1\}$ be the set of classical bits and let $\{\drc{0},\drc{1}\}$ denote the canonical basis of $\hache_1$.

For any $n\geq 1$ and any $\vcg{\varepsilon} = \varepsilon_{n-1}\cdots\varepsilon_1\varepsilon_0\in Q^n$ let
$\drc{\vcg{\varepsilon}} = \drc{\varepsilon_{n-1}}\otimes \cdots\otimes\drc{\varepsilon_1}\otimes\drc{\varepsilon_0}.$
Then $\left(\drc{\vcg{\varepsilon}}\right)_{\vcgs{\varepsilon}\in Q^n}$ is the canonical basis of $\hache_n$.

For any $n\geq 2$ and any $\vcg{\varepsilon} = \varepsilon_{n-1}\cdots\varepsilon_1\varepsilon_0\in Q^n$ let
${\bf b}_{\vcgs{\varepsilon}} = \frac{1}{\sqrt{2}}\left(\drc{0\varepsilon_{n-2}\cdots\varepsilon_1\varepsilon_0} + 
(-1)^{\varepsilon_{n-1}}\drc{1\overline{\varepsilon_{n-2}}\cdots\overline{\varepsilon_1}\,\overline{\varepsilon_0}}\right),$
where $\overline{\varepsilon_i}$ is the orthogonal complement element of $\varepsilon_i$, for each $i=n-2,\ldots,0$. 

Then $\left({\bf b}_{\vcgs{\varepsilon}}\right)_{\vcgs{\varepsilon}\in Q^n}$ is the Bell basis of $\hache_n$, and it consists of maximally entangled states.

The group $\mbox{U}(2^n)$ consists of all unitary linear transforms $\hache_n\to\hache_n$, with map composition as operation, and $\mbox{\rm SU}(2^n)$ is the subgroup of orientation preserving unitary transforms: $\forall T\in\mbox{\rm SU}(2^n)$, $\det T = 1$. 


Let $S_1(\C)$ the unit sphere of $\hache_1 = \C^2$,
$S_1(\C) = \{{\bf x}\in\hache_1|\ {\bf x}^H{\bf x}=1\}$ is the set of {\em qubits}.
The tensor square power $\hache_2 =  \hache_1 \otimes   \hache_1$ is the four-dimensional complex Hilbert space. Its 
unit sphere $S_3(\C) = \{{\bf x}\in\hache_2|\ {\bf x}^H{\bf x}=1\}$ is the set of 2-{\em quregisters}. In 
this case, ${\cal L}(\hache_2)\approx\C^{2^2\times 2^2}$ is the space of square $(2^2\times 2^2)$-complex matrices and $\mbox{\rm D}(\hache_2)$ is the set of positive matrices with trace 1.
For any 2-quregister ${\bf x}=(x_0,x_1,x_2,x_3)\in S_3(\C)$ we have
$$\rho({\bf x}) = {\bf x}{\bf x}^H = \left[
\begin{array}{cccc}
 \left| x_0\right| ^2 & x_0 \overline{x_1} & x_0 \overline{x_2} & x_0 \overline{x_3} \\
 x_1 \overline{x_0} & \left| x_1\right| ^2 & x_1 \overline{x_2} & x_1 \overline{x_3} \\
 x_2 \overline{x_0} & x_2 \overline{x_1} & \left| x_2\right| ^2 & x_2 \overline{x_3} \\
 x_3 \overline{x_0} & x_3 \overline{x_1} & x_3 \overline{x_2} & \left| x_3\right| ^2 \\
\end{array}
\right].$$
The reduced trace matrices are 
$$\tri{0}{\rho({\bf x})} = \left[
\begin{array}{cc}
 \left| x_0\right| ^2+\left| x_1\right| ^2 & x_0 \overline{x_2}+x_1 \overline{x_3} \\
 x_2 \overline{x_0}+x_3 \overline{x_1} & \left| x_2\right| ^2+\left| x_3\right| ^2 \\
\end{array}
\right]\ ,\ 
\tri{1}{\rho({\bf x})} = \left[
\begin{array}{cc}
 \left| x_0\right| ^2+\left| x_2\right| ^2 & x_0 \overline{x_1}+x_2 \overline{x_3} \\
 x_1 \overline{x_0}+x_3 \overline{x_2} & \left| x_1\right| ^2+\left| x_3\right| ^2 \\
\end{array}
\right].$$
The reduced trace matrices have the same eigenvalues. Those are
\begin{equation}
\lambda_0 = \frac{1}{2} \left(1-s({\bf x})\right)\ \ ,\ \ \lambda_1 = \frac{1}{2} \left(1+ s({\bf x})\right) \label{eq.ells}
\end{equation}
where
\begin{equation}
s({\bf x}) = \sqrt{1-4 |t({\bf x})|^2}\ \mbox{ and }\ t({\bf x}) = x_0 x_3-x_1 x_2, \label{eq.ells1}
\end{equation}
Hence, the Von Neumann entropy of the reduced trace matrices is
\begin{eqnarray*}
E(\rho({\bf x})) &=& - \lambda_0\log_2\lambda_0 -  \lambda_1\log_2\lambda_1 \\
 &=& -\frac{1}{2} \left(1-s({\bf x})\right)\left(\log_2\left(1-s({\bf x})\right) - 1\right) -  \frac{1}{2} \left(1+s({\bf x})\right)\left(\log_2\left(1+s({\bf x})\right) - 1\right) \\
 &=& -\frac{1}{2} \left(\log_2\left( \left(1-s({\bf x})\right) \left(1+s({\bf x})\right)\right) + s({\bf x})\log_2\left(\frac{1+s({\bf x})}{1-s({\bf x})}\right)\right) +1 \\
 &=& -\frac{1}{2} \left(\log_2\left( 1-s({\bf x})^2\right) + s({\bf x})\log_2\left(\frac{1-s({\bf x})^2}{(1-s({\bf x}))^2}\right)\right) +1 \\
 &=& -\frac{1}{2} \left( 1+s({\bf x})\right)\log_2\left( 1-s({\bf x})^2\right) +  s({\bf x})\log_2\left(1-s({\bf x})\right) +1 
\end{eqnarray*}
which is a measure of the entanglement of $\rho({\bf x})$ in $\mbox{\rm D}(\hache_2)$. Detailed definitions and constructions
of the above results are shown in  Section \ref{Dm}.    

On the other hand, from~(\ref{eq.ells}) we see that $\lambda_0+\lambda_1=1$, namely $\lambda_1= 1-\lambda_0$ and 
\begin{eqnarray}
E(\rho({\bf x})) &=& -\lambda_0\log_2\lambda_0 -  (1-\lambda_0)\log_2(1-\lambda_0)  \nonumber \\
 &=& -\frac{1}{2} \left( \left( 1-s({\bf x})\right)\log_2\left( 1-s({\bf x})\right) +  \left( 1+s({\bf x})\right)\log_2\left( 1+s({\bf x})\right)\right) +1  \label{eq.olm} .
\end{eqnarray}
Thus, 
\begin{equation}
E(\rho({\bf x})) =  0\ \Longleftrightarrow\ \lambda_0\in\{0,1\}\ \Longleftrightarrow\ s({\bf x})=\sqrt{1}\ \Longleftrightarrow\ t({\bf x})= 0\ \Longleftrightarrow\ x_1 x_2=x_0 x_3, \label{eq.eqs}
\end{equation}
and this last condition entails that ${\bf x}\in S_3(\C)$ is a separable unit vector. These separability criteria 
are included in the Proposition \ref{pr.sp}.


\section{Case of qubits}

\subsection{Embedding $S_1(\C)$ into $\mbox{\rm SU}(2)$}

Let us define 
\begin{equation}
\Psi_1: S_1(\C)\to\mbox{\rm SU}(2)\ \ ,\ \ {\bf x}=\left[\begin{array}{c}
 x_0 \\ x_1 
\end{array}\right] \mapsto \Psi_1({\bf x}) = \left[\begin{array}{rr}
 x_0 & -\overline{x_1} \\
 x_1 & \overline{x_0}
 \end{array}\right].
 \label{eq.psi1}
\end{equation}
Via the map $\Psi_1$, any qubit is identified with an element of $\mbox{\rm SU}(2)$. Conversely, 
if $X = \left[\begin{array}{rr}
 x_0 & y_0 \\
 x_1 & y_1
 \end{array}\right]\in\mbox{\rm SU}(2)$, then 
 $x_0y_1-x_1y_0=1.$
By assuming $(x_0,x_1)\in S_1(\C)$, the solutions of this last equation are the points $(y_0,y_1)\in\hache_1$ in the straight line passing through $(-\overline{x_1},\overline{x_0})$ parallel to the straight-line orthogonal to $(x_0,-x_1)$. Since $(x_0,-x_1)\in S_1(\C)$, this line is tangent to $S_1(\C)$ at this point. The solution line is parameterised thus as
 $Y(y) = (-\overline{x_1},\overline{x_0}) + y(1,-\overline{x_0^{-1}x_1})$, with $y\in\C.$
 Also $(-\overline{x_1},\overline{x_0})\in S_1(\C)$, and the only solution $(y_0,y_1)$ of  $x_0y_1-x_1y_0=1$ in $S_1(\C)$ is $(y_0,y_1) =  (-\overline{x_1},\overline{x_0})$. Thus $X=\Psi_1(x_0,x_1)$. Hence, $\Psi_1$ is a bijection $S_1(\C)\to\mbox{\rm SU}(2)$. 
 
 The operation in the group $\mbox{\rm SU}(2)$ translated into $S_1(\C)$ is
 \begin{equation}
\star_1:S_1(\C)\times S_1(\C)\to S_1(\C)\ \ ,\ \ 
 \left( \left[\begin{array}{c} x_{00} \\ x_{10} \end{array}\right] , 
 \left[\begin{array}{c} x_{01} \\ x_{11} \end{array}\right]\right)
 \mapsto
  \left[\begin{array}{c} x_{00} \\ x_{10} \end{array}\right] \star_1 
 \left[\begin{array}{c} x_{01} \\ x_{11} \end{array}\right] =
 \left[\begin{array}{c} x_{00}x_{01} -\overline{x_{10}} x_{11} \\ x_{10}x_{01} + \overline{x_{00}} x_{11} \end{array}\right],
  \label{eq.psi2}
\end{equation}
hence $(S_1(\C),\star_1,\left[\begin{array}{c} 1 \\ 0 \end{array}\right])$ is a group. 

In fact, for a qubit ${\bf x} = [x_0\ \ x_1]^T\in S_1(\C)$ we have
\begin{eqnarray*}
{\bf x}^{\star_1 1} &=& \left[\begin{array}{c} x_0 \\ x_1 \end{array}\right], \\
{\bf x}^{\star_1 2} &=& \left[\begin{array}{c} x_0^2-x_1 \overline{x_1} \\ 2 x_1 \Re(x_0) \end{array}\right],\\
{\bf x}^{\star_1 3} &=&\left[\begin{array}{c} x_0^3-x_1 (2 \Re(x_0)+x_0) \overline{x_1} \\ x_1 \left(\left(\overline{x_0}\right)^2+2 x_0 \Re(x_0)-x_1 \overline{x_1}\right) \end{array}\right],\\
{\bf x}^{\star_1 4} &=&\left[\begin{array}{c} \left(x_0^2-x_1 \overline{x_1}\right)^2-4 x_1 \Re(x_0)^2 \overline{x_1} \\ x_1 \left(x_0^3+x_0 \left|x_0\right| ^2+2
   \Re(x_0) \left(\left(\overline{x_0}\right)^2-2 x_1 \overline{x_1}\right)\right) \end{array}\right],
\end{eqnarray*}
where $\Re(z)$ denotes the real part of the complex number $z\in\C$.

We recall that the {\em order} of an element ${\bf x}\in S_1(\C)$ is 
$o({\bf x}) =  \min_{n\in\N} \left\{{\bf x}^{\star_1 n} = [1\ \ 0]^T\right\}$. For instance,
$$o\left(\left[\begin{array}{c} 1 \\ 0 \end{array}\right]\right) = 1\ \ ,\ \ o\left(\left[\begin{array}{r} -1 \\ 0 \end{array}\right]\right) = 2\ \ ,\ \ o\left(\left[\begin{array}{c} 0 \\ 1 \end{array}\right]\right) = 4\ \ ,\ \ o\left(\left[\begin{array}{r} 0 \\ -1 \end{array}\right]\right) = 4,$$
while
$$\forall \varepsilon_0,\varepsilon_1\in\{-1,+1\}:\ o\left(\frac{1}{\sqrt{2}}\left[\begin{array}{c} \varepsilon_0 \\ \varepsilon_1 \end{array}\right]\right) = 8.$$

As a direct consequence of the {\em Poincar\'e's Recurrence Theorem}~\cite{Stein05} we have:
\begin{prop}
 Let $r_0,r_1\in[0,1]$ be such that $|r_0|^2+|r_1|^2=1$ and $t_0,t_1\in\R$ two irrational numbers. Let ${\bf x} = [r_0\,\mbox{\rm exp}\left(\imb\, t_0\right)\ \ r_1\,\mbox{\rm exp}\left(\imb\, t_1\right)]^T$, where $\imb = \sqrt{-1}$. Then the subgroup $\left\langle{\bf x}\right\rangle = \{{\bf x}^{\star_1 n}|\ n\in\Z\}<S_1(\C)$, generated by ${\bf x}$, is a countable dense subgroup of $S_1(\C)$.
\end{prop}

We recall that a {\em quantum gate} $U$ is a unitary map $U:\hache_1\to\hache_1$, namely $U^HU=\id{2}$. 
Thus $U\in\mbox{U}(2)$ and it is a bijection $S_1(\C)\to S_1(\C)$ when restricted to the unit sphere $S_1(\C)$.

A mechanical computation suffices to prove the following:
\begin{prop} \label{pr.cm1} A quantum gate $U$ commutes with the bijection 
$\Psi_1:S_1(\C)\to\mbox{\rm SU}(2)$ if and only if $\det U = 1$. In symbols:
$$\forall U\in\mbox{U}(2):\ \left[\Psi_1\circ U = U\circ\Psi_1\ \Longleftrightarrow\ U\in\mbox{\rm SU}(2)\right],$$
where $\circ$ is the composition of maps.
\end{prop}

\section{Case of 2-quregisters}

\subsection{Embedding $S_3(\C)$ into $\mbox{SL}(2^2)$} \label{sc.n1}

Let ${\bf c}_0 = [c_{00}\ \ c_{10}]^T$, ${\bf c}_1 = [c_{01}\ \ c_{11}]^T$ be two qubits in the unit sphere $S_1(\C)$ of the Hilbert space $\hache_1$ and let $\Psi_1: S_1(\C)\to\mbox{\rm SU}(2)$ be the bijection defined as in~(\ref{eq.psi1}). Then,
\begin{equation}
\Psi_1({\bf c}_0)\otimes\Psi_1({\bf c}_1) = \left[
\begin{array}{rrrr}
 c_{00} c_{01} & -c_{00} \overline{c_{11}} & -\overline{c_{10}}c_{01}  &  \overline{c_{10}} \overline{c_{11}} \\
 c_{00} c_{11} & c_{00} \overline{c_{01}} & - \overline{c_{10}}c_{11} &  -\overline{c_{10}} \overline{c_{01}} \\
 c_{10} c_{01} & -c_{10} \overline{c_{11}} & \overline{c_{00}}c_{01}  & -\overline{c_{00}} \overline{c_{11}} \\
 c_{10} c_{11} & c_{10} \overline{c_{01}} & \overline{c_{00}}c_{11}  & \overline{c_{00}}  \overline{c_{01}} 
\end{array}
\right]. \label{eq.ps3}
\end{equation}

The collection of 2-quregisters is the unit sphere $S_3(\C)$ of the Hilbert space $\hache_2$. A natural embedding of the Cartesian product $S_1(\C)\times S_1(\C)$ into $S_3(\C)$ is given by the injective map
\begin{equation}
I_2:S_1(\C)\times S_1(\C)\to S_3(\C)\ \ ,\ \ ({\bf c}_0,{\bf c}_1)\mapsto I_2({\bf c}_0,{\bf c}_1) = {\bf c}_0\otimes{\bf c}_1. \label{eq.ps31}
\end{equation}
A 2-quregister ${\bf x}\in S_3(\C)$ is {\em separable} if it is in the image of $I_2$, namely, there exist ${\bf c}_0,{\bf c}_1\in S_1(\C)$ such that ${\bf x} = {\bf c}_0\otimes{\bf c}_1$.

\begin{prop} \label{pr.sp} A $2$-quregister ${\bf x} = [x_0\ \ x_1\ \ x_2\ \ x_3]^T\in S_3(\C)$ is separable if and only if 
$x_0x_3 = x_1x_2.$
\end{prop}

Let $\mbox{\it Sp}_2\subset S_3(\C)$ be the collection of $2$-quregisters that are separable. Thus, the condition 
at Proposition~\ref{pr.sp}  is a defining predicate of the set $\mbox{\it Sp}_2$.

For each index $j\in\{0,1,2,3\}$, let $C_j = 
\{{\bf x}\in S_3(\C)|\ x_j\not=0\} = S_3(\C)\cap\pi_j(\C-\{0\})$, where $\pi_j$ is the $j$-th canonical projection.
Each set $C_j$ is an open set in the unit sphere $S_3(\C)$, with the topology induced by the 
Hilbert space $\hache_2$, and they cover $S_3(\C)$. Namely, $\left(C_j\right)_{j=0}^3$ is an open 
covering of $S_3(\C)$ and it determines a structure of a complex 3-dimensional differential manifold 
in $S_3(\C)$: each set $C_j$ is a 3-dimensional complex chart in $S_3(\C)$.

\begin{prop}\label{pr.sep} If a $2$-quregister ${\bf x} = [x_0\ \ x_1\ \ x_2\ \ x_3]^T\in S_3(\C)$ is separable, 
then it can be tensor splited as ${\bf x} = {\bf c}_0\otimes{\bf c}_1$, where the qubits ${\bf c}_0,{\bf c}_1$ are 
determined according to the following rules:
\begin{eqnarray*}
{\bf x}\in C_0 &\Longrightarrow& 
{\bf c}_0 = \frac{1}{r_{02}}\left[\begin{array}{c}
 x_0 \vspace{2ex} \\ x_2
\end{array}\right]\ \land\ 
{\bf c}_1 = r_{02} \left[\begin{array}{c}
 1 \vspace{2ex} \\ \frac{x_1}{x_0}
\end{array}\right] \mbox{ with } r_{02} = \sqrt{|x_0|^2+|x_2|^2}, \\
{\bf x}\in C_1 &\Longrightarrow& 
{\bf c}_0 = \frac{1}{r_{13}}\left[\begin{array}{c}
 x_1 \vspace{2ex} \\ x_3
\end{array}\right]\ \land\ 
{\bf c}_1 = r_{13} \left[\begin{array}{c}
 \frac{x_0}{x_1} \vspace{2ex} \\ 1
\end{array}\right] \mbox{ with } r_{13} = \sqrt{|x_1|^2+|x_3|^2}, \\
{\bf x}\in C_2 &\Longrightarrow& 
{\bf c}_0 = \frac{1}{r_{02}}\left[\begin{array}{c}
 x_0 \vspace{2ex} \\ x_2
\end{array}\right]\ \land\ 
{\bf c}_1 = r_{02} \left[\begin{array}{c}
 1 \vspace{2ex} \\ \frac{x_3}{x_2}
\end{array}\right], \\ 
{\bf x}\in C_3 &\Longrightarrow& 
{\bf c}_0 = \frac{1}{r_{13}}\left[\begin{array}{c}
 x_1 \vspace{2ex} \\ x_3
\end{array}\right]\ \land\ 
{\bf c}_1 = r_{13} \left[\begin{array}{c}
 \frac{x_2}{x_3} \vspace{2ex} \\ 1
\end{array}\right]. 
\end{eqnarray*}
Besides since ${\bf x} = {\bf c}_0\otimes{\bf c}_1$, for any unit complex number $u\in\C$, ${\bf x} = (u^{-1}{\bf c}_0)\otimes (u{\bf c}_1)$ is another tensor split of ${\bf x}$.
\end{prop}

For $k\in\{0,1,2,3\}$ and a unit complex number $u\in\C$, let $\Phi_{2ku}:C_k\to\C^{2^2\times 2^2}$ be defined as follows, with ${\bf x}\in C_k$:

\newcommand{\eb}{\hspace{.3cm}}

\begin{eqnarray}
\Phi_{20u}({\bf x}) &=& \left[
\begin{array}{rrrr}
 x_0 & -u^{-2}\xi(x_0)^2\overline{x_1}  & -u^2 \overline{x_2} & \overline{x_3} \\
 x_1 & u^{-2}x_0 & -u^2\xi(x_2)^{-2} x_3 & -\overline{x_2} \\
 x_2 & -u^{-2} \xi\left(\frac{x_1}{x_0}\right)^{-2}x_3 & u^2 \overline{x_0} & -\overline{x_1}   \\
 x_3 & u^{-2}x_2 & u^2\xi(x_0)^{-2} x_1 & \overline{x_0} 
\end{array}
\right]
,\label{eq.ps330} \\
\Phi_{21u}({\bf x}) &=& \left[
\begin{array}{rrrr}
 x_0 & -u^{-2}x_1 & -u^2\xi(x_3)^{-2} x_2 & \overline{x_3} \\
 x_1 & u^{-2}\xi(x_1)^2 \overline{x_0} & -u^2 \overline{x_3} & -\overline{x_2} \\
 x_2 & -u^{-2}x_3 & u^2\xi(x_1)^{-2}  x_0 & -\overline{x_1} \\
 x_3 & u^{-2} \xi\left(\frac{x_0}{x_1}\right)^{-2}x_2 & u^2 \overline{x_1} & \overline{x_0} 
\end{array}
\right]
,\label{eq.ps331}  \\
\Phi_{22u}({\bf x}) &=& \left[
\begin{array}{rrrr}
 x_0 & -u^{-2} \xi\left(\frac{x_3}{x_2}\right)^{-2}x_1 & -u^2 \overline{x_2} & \overline{x_3}   \\
 x_1 & u^{-2}x_0 & -u^2\xi(x_2)^{-2}  x_3 & -\overline{x_2} \\
 x_2 & -u^{-2}\xi(x_2)^2\overline{x_3}  & u^2 \overline{x_0} & -\overline{x_1} \\
 x_3 & u^{-2}x_2 & u^2\xi(x_0)^{-2}  x_1 & \overline{x_0} 
\end{array}
\right]
,\label{eq.ps332} \\
\Phi_{23u}({\bf x}) &=& \left[
\begin{array}{rrrr}
 x_0 & -u^{-2}x_1 & -u^2\xi(x_3)^{-2} x_2 & \overline{x_3} \\
 x_1 & u^{-2} \xi\left(\frac{x_2}{x_3}\right)^{-2}x_0 & -u^2 \overline{x_3} & -\overline{x_2}   \\
 x_2 & -u^{-2}x_3 & u^2\xi(x_1)^{-2} x_0 & -\overline{x_1} \\
 x_3 & u^{-2}\xi(x_3)^2\overline{x_2} & u^2 \overline{x_1} & \overline{x_0} 
\end{array}
\right] 
,\label{eq.ps333}
\end{eqnarray}
where $\xi:\C\to\C$ is the map that for any non-zero complex number takes the unit complex number along its own direction,
$$z\mapsto\xi(z) = \left\{ \begin{array}{cl}
 \frac{z}{|z|} & \mbox{ if }z\not = 0, \\
 1 & \mbox{ if }z = 0. 
\end{array}\right.$$
Direct computations show that for any separable $2$-quregister ${\bf x}\in \mbox{\it Sp}_2$, since the 
condition at Proposition~\ref{pr.sp} holds, for any $k\in\{0,1,2,3\}$, 
$\Phi_{2ku}({\bf x})^H\Phi_{2ku}({\bf x})=\id{2^2}.$
Consequently for $k\in\{0,1,2,3\}$, the map
$\Phi_{2ku}:C_k\to\C^{2^2\times 2^2}$
is such that it determines an embedding of $C_k\cap\mbox{\it Sp}_2$ into the symmetry group $\mbox{\rm SU}(2^2)$.

In the case of a separable 2-quregister ${\bf x} \in C_k\cap\mbox{\it Sp}_2$, for a tensor split ${\bf x} = {\bf c}_0\otimes{\bf c}_1$ we have that the matrix $\Psi_1({\bf c}_0)\otimes\Psi_1({\bf c}_1)$ at~(\ref{eq.ps3}) coincides with $\Phi_{23u}({\bf x})$ being $u=\xi(c_k)$ the unit complex number in the direction of the complex number $c_k\in\C$, where $c_0=c_{00}$, $c_1=c_{10}$, $c_2=c_{01}$, and $c_3=c_{11}$.
\bigskip

\noindent{\bf Example \ref{sc.n1}.1.} The $i$-th vector ${\bf e}_i = \left[\delta_{ij}\right]_{j=0}^3$ in the canonical basis of $\hache_2$ is a separable $2$-quregister:
$${\bf e}_{2i_1+i_0} = {\bf e}_{i_1}\otimes{\bf e}_{i_0}.$$ 
We have ${\bf e}_i\in C_i$ while ${\bf e}_i\not\in C_j$, for $j\not=i$. Then for any unit complex number $u\in\C$:
\begin{equation}
\begin{array}{ll}
 \Phi_{20u}({\bf e}_0) = \left[
\begin{array}{cccc}
 1 & 0 & 0 & 0 \\
 0 & u^{-2} & 0 & 0 \\
 0 & 0 & u^2 & 0 \\
 0 & 0 & 0 & 1 \\
\end{array}
\right], &
 \Phi_{21u}({\bf e}_1) = \left[
\begin{array}{cccc}
 0 & -u^{-2} & 0 & 0 \\
 1 & 0 & 0 & 0 \\
 0 & 0 & 0 & -1 \\
 0 & 0 & u^2 & 0 \\
\end{array}
\right], \vspace{2ex} \\
 \Phi_{22u}({\bf e}_2) = \left[
\begin{array}{cccc}
 0 & 0 & -u^2 & 0 \\
 0 & 0 & 0 & -1 \\
 1 & 0 & 0 & 0 \\
 0 & u^{-2} & 0 & 0 \\
\end{array}
\right], &
 \Phi_{23u}({\bf e}_3) = \left[
\begin{array}{cccc}
 0 & 0 & 0 & 1 \\
 0 & 0 & -u^2 & 0 \\
 0 & -u^{-2} & 0 & 0 \\
 1 & 0 & 0 & 0 \\
\end{array}
\right]. 
\end{array} \label{eq.ex01}
\end{equation}
Being separable the vectors at the canonical basis, all matrices above are unitary.
\bigskip

\noindent{\bf Example \ref{sc.n1}.2.} The $i$-th vector ${\bf b}_i$ in the Bell basis of $\hache_2$ is a maximally entangled $2$-quregister:
$${\bf b}_{2i_1+i_0} = \frac{1}{\sqrt{2}}\left({\bf e}_{0}
\otimes{\bf e}_{i_1}+(-1)^{i_0}{\bf e}_{1}\otimes{\bf e}_{1+i_1}\right).$$
Each $2$-quregister ${\bf b}_i$ is in two charts $C_k$. For any unit complex number $u\in\C$:
$$\begin{array}{ll}
 \Phi_{20u}({\bf b}_0) =  \Phi_{23u}({\bf b}_0) = \frac{1}{\sqrt{2}} \left[
\begin{array}{cccc}
 1 & 0 & 0 & 1 \\
 0 & u^{-2} & -u^2 & 0 \\
 0 & -u^{-2} & u^2 & 0 \\
 1 & 0 & 0 & 1 \\
\end{array}
\right], &
 \Phi_{20u}({\bf b}_1) = \Phi_{23u}({\bf b}_1) =  \frac{1}{\sqrt{2}}\left[
\begin{array}{cccc}
 1 & 0 & 0 & -1 \\
 0 & u^{-2} & u^2 & 0 \\
 0 & u^{-2} & u^2 & 0 \\
 -1 & 0 & 0 & 1 \\
\end{array}
\right], \vspace{2ex} \\
 \Phi_{21u}({\bf b}_2) = \Phi_{22u}({\bf b}_2) = \frac{1}{\sqrt{2}} \left[
\begin{array}{cccc}
 0 & -u^{-2} & -u^2 & 0 \\
 1 & 0 & 0 & -1 \\
 1 & 0 & 0 & -1 \\
 0 & u^{-2} & u^2 & 0 \\
\end{array}
\right], &
 \Phi_{21u}({\bf b}_3) = \Phi_{22u}({\bf b}_3) = \frac{1}{\sqrt{2}} \left[
\begin{array}{cccc}
 0 & -u^{-2} & u^2 & 0 \\
 1 & 0 & 0 & 1 \\
 -1 & 0 & 0 & -1 \\
 0 & -u^{-2} & u^2 & 0 \\
\end{array}
\right]. 
 \end{array}$$
 No above matrix is unitary. In fact, if $B$ is any of the above matrices, then $B^HB$ has 2 and 0 as eigenvalues, each of multiplicity 2. Hence, the spectral norm of $B$ is $\|B\|_2 = \sqrt{2}$.
\bigskip

\noindent{\bf Example \ref{sc.n1}.3.} Consider a vector ${\bf x}_p =(\sqrt{p},0,0,\sqrt{1-p}) = \sqrt{p}\,{\bf e}_0 + \sqrt{1-p}\,{\bf e}_3\in S_3(\C)$, with $p\in[0,1]$. Then ${\bf x}_p\in C_0\cap C_3$. For any unit complex number $u\in\C$ we have
$$\Phi_{20u}({\bf x}_p)  = \Phi_{23u}({\bf x}_p) = \left[
\begin{array}{cccc}
 \sqrt{p} & 0 & 0 & \sqrt{1-p} \\
 0 & u^{-2}\sqrt{p} & -u^2 \sqrt{1-p} & 0 \\
 0 & -u^{-2}\sqrt{1-p} & u^2 \sqrt{p} & 0 \\
 \sqrt{1-p} & 0 & 0 & \sqrt{p} \\
\end{array}
\right]. $$
In fact, from relations~(\ref{eq.ex01}), we have
$$\Phi_{20u}({\bf x}_p)  =  \sqrt{p}\,\Phi_{20u}({\bf e}_0) +  \sqrt{1-p}\,\Phi_{23u}({\bf e}_3).$$
Then, for the matrix
$$V = \left[
\begin{array}{cccc}
 1 & 0 & 0 & 2 \sqrt{1-p} \sqrt{p} \\
 0 & 1 & -2 \sqrt{1-p} \sqrt{p} & 0 \\
 0 & -2 \sqrt{1-p} \sqrt{p} & 1 & 0 \\
 2 \sqrt{1-p} \sqrt{p} & 0 & 0 & 1 \\
\end{array}
\right],$$
we have $\Phi_{20u}({\bf x}_p)^H\Phi_{20u}({\bf x}_p) = V = \Phi_{23u}({\bf x}_p)^H\Phi_{23u}({\bf x}_p)$, and the eigenvalues of $V$ are $1 - 2 \sqrt{(1 - p) p}$, $1 + 2 \sqrt{(1 - p) p}$, each with characteristic 2. Hence, the spectral norm is $\|\Phi_{20u}({\bf x}_p)\|_2 = \sqrt{1 + 2 \sqrt{(1 - p) p}}$.

We observe that for $p=0$, ${\bf x}_0 = {\bf e}_3$, the fourth vector in the canonical basis, which is separable, for $p=\frac{1}{2}$, ${\bf x}_{\frac{1}{2}} = {\bf b}_0$, the first vector in the Bell basis, which is maximally entangled and for $p=1$, ${\bf x}_1 = {\bf e}_0$, the first vector in the canonical basis, which is separable.

\begin{figure}[!t]
\centering
\includegraphics[width=4in]{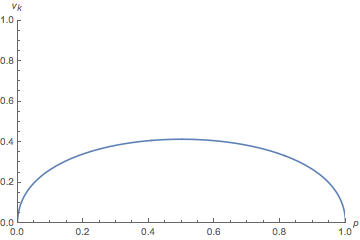} \hspace{1em} 
\caption{Graph of the spectral norm of the transforms of the 2-quregister ${\bf x}_p$ translated by $-1$.}
\label{fg.01}
\end{figure}

In Figure~\ref{fg.01} it is displayed the plot of the map
$p\mapsto\|\Phi_{20u}({\bf x}_p)\|_2-1 = \sqrt{1 + 2 \sqrt{(1 - p) p}}-1.$
\bigskip

For any 2-quregister ${\bf x}\in C_k$, let $\nu_k({\bf x}) = \|\Phi_{2ku}({\bf x}_p)\|_2-1$, with $u\in\C$ being a unit complex number. 
 
As a characterisation of the set $\mbox{\it Sp}_2\subset S_3(\C)$ of separable 2-quregisters, we have:
 
 \begin{prop}\label{pr.seep} For any $k\in\{0,1,2,3\}$, any unitary complex number $u\in \C$ and any ${\bf x}\in C_k$:
 $${\bf x}\in \mbox{\it Sp}_2\ \Longleftrightarrow\ \Phi_{2ku}({\bf x})\in\mbox{\it SU}(2^2)\ \Longleftrightarrow\ \nu_k({\bf x}) = 0.$$
 \end{prop}

The map $\nu_k$ can be considered a measure of entanglement. It satisfies the conventional conditions of an entanglement measure~\cite{Don02}:
\begin{description}
\item[Separability. ] If ${\bf x}\in S_3(\C)$ is separable, then $\nu_k({\bf x})=0$.
\item[Normality. ] In the maximally entangled vectors, $\nu_k$ attains its maxima. Indeed, the measure $\frac{2}{\sqrt{2}-1}\nu_k$ has $2=\log_22^2$ as maximal value.
\item[Continuity. ] $\nu_k$ is continuous with respect to the topology of $S_3(\C)$.
\item[Boundedness under local operations. ] The entanglement cannot be increased by applying local operations.
\end{description}
Let us check this last assertion.

%
For each $k\in\{0,1,2,3\}$ and a unit complex number $u\in\C$, let $\Phi_{2ku}:C_k\to\C^{2^2\times 2^2}$ be defined by relations~(\ref{eq.ps330}, \ref{eq.ps331}, \ref{eq.ps332}, \ref{eq.ps333}) respectively.

First let us state as a proposition the following result, which can be proved in and exhaustive way through direct calculations.

\begin{prop} For each $k\in\{0,1,2,3\}$ and each ${\bf x}\in C_k$ there is a unit complex number $u=u({\bf x})\in\C$, such that
$$\Phi_{2ku}^H({\bf x})\,\Phi_{2ku}({\bf x})\left[
\begin{array}{cccc}
 1 & 0 & 0 & 2\, \overline{t({\bf x})} \\
 0 & 1 & 0 & 0 \\
 0 & 0 & 1 & 0 \\
 2\, t({\bf x}) & 0 & 0 & 1 
\end{array}
\right]
= Z_u({\bf x})$$
where $t({\bf x})$ is defined by~(\ref{eq.ells1}),
\end{prop}

\begin{prop} The spectrum of the matrix $Z_u({\bf x})$ is
$\Lambda_u({\bf x})=\{1,1,1-2 \left| t({\bf x})\right| ,1+2 \left| t({\bf x})\right|\},$ and the corresponding eigenvectors are the columns of the matrix
$$W_u({\bf x}) = \left[
\begin{array}{cccc}
 0 & 0 & -\frac{\overline{t({\bf x})}}{\left| t({\bf x})\right| } & 1 \\
 0 & 1 & 0 & 0 \\
 1 & 0 & 0 & 0 \\
 0 & 0 & 1 & \frac{\overline{t({\bf x})}}{\left| t({\bf x})\right| }
\end{array}
\right].$$
\end{prop}

Hence the introduced measure $\nu_k$ is such that  
\begin{equation}
\nu_k({\bf x}) = \sqrt{1+2 \left| t({\bf x})\right|}-1. \label{eq.nud}
\end{equation}
Let us compare roughly $\nu_k({\bf x})$ as in~(\ref{eq.nud}) with the Von Neumann entropy $E(\rho({\bf x}))$ as in~(\ref{eq.olm}). According
to Proposition~\ref{rm.spsp} and the equivalences in~(\ref{eq.eqs}):
$${\bf x}\mbox{ is separable }\ \Longleftrightarrow\ E(\rho({\bf x}))=0\ \Longleftrightarrow\ t({\bf x})= 0\ \Longleftrightarrow\ \nu_k({\bf x}).$$
On the other hand,
\begin{itemize}
\item according to~(\ref{eq.nud}), maximal values of $\nu_k({\bf x})$ correspond to maximal values of $t({\bf x})$,
\item according to~(\ref{eq.ells1}), if $t({\bf x})=\frac{1}{2}$, then $s({\bf x})=0$, $\lambda_0=\frac{1}{2}$ and both $E(\rho({\bf x}))=1$ and $\nu_k({\bf x})=\sqrt{2}-1$ attain their maximum values, and
\item according to~(\ref{eq.ells1}), if $|t({\bf x})|>\frac{1}{2}$, then $s({\bf x}),\lambda_0\in\C-\R$, thus $E(\rho({\bf x}))$ and $\nu_k({\bf x})$ cannot further be compared.
\end{itemize}

\begin{prop} For any linear $U\in{\cal L}(\hache_1)$ and any unitary vector ${\bf x}\in S_3(\C)$,
$$t((U\otimes\id{2}){\bf x}) = (\det U) t({\bf x}) = t((\id{2}\otimes U){\bf x}).$$
Hence, if $U$ is unitary, then
$$\left|t((U\otimes\id{2}){\bf x})\right| = \left|t({\bf x})\right| = \left|t((\id{2}\otimes U){\bf x})\right|.$$
\end{prop}

\begin{prop} The introduced measure, $\nu_k$ is not increasing under the application of local operators with 
classical communication (LOCC), or stochastic LOCC (SLOCC).
\end{prop}




 Finally, since the spectral bound of a matrix $M=\left(m_{ij}\right)_{i,j\in\{0,1,2,3\}}\in\C^{2^2\times 2^2}$ is bounded as
 $$\|M\|_2 \leq \left(\sum_{i,j=0}^3|m_{ij}|^2\right)^{\frac{1}{2}},$$
 from relations~(\ref{eq.ps330})--(\ref{eq.ps333}) we have
 $$\forall k\in\{0,1,2,3\},u\in\C\mbox{ with }|u|=1,{\bf x}\in C_k:\ \ \|\Phi_{2ku}({\bf x})\|_2 \leq 2,$$
but this is not a tight bound.

\subsection{Density matrices}\label{Dm}

Let ${\bf x} = (x_0,x_1,x_2,x_3)\in S_3(\C)$ be a 2-quregister. The {\em projection} along 
the direction of this vector is $\pi_{\bf x}:\hache_2\to\hache_2$, ${\bf y}\mapsto \pi_{\bf x}({\bf y}) = 
({\bf x}\,{\bf x}^H){\bf y}$. The matrix $\rho_{2}({\bf x}) = {\bf x}\,{\bf x}^H$ is the {\em density matrix} 
determined by ${\bf x}$.The notations $\rho_2$ and $\rho_1$ refer to given in the Preliminaries (\ref{Preliminaries}),
where the subindices emphasize the domains $\hache_2$ and $\hache_1$ respectively of the proyections.

Similarly for a qubit ${\bf z}\in S_1(\C)$, its density matrix is $\rho_{1}({\bf z}) = {\bf z}\,{\bf z}^H$.

A mechanical computation suffices to prove the following:
\begin{prop} \label{pr.cdm} If a separable 2-quregister is factored as in Proposition~\ref{pr.sep}, say 
${\bf x} = {\bf c}_0\otimes{\bf c}_1$, then
$$\rho_{2}({\bf x}) = \rho_{1}({\bf c}_0)\otimes\rho_{1}({\bf c}_1).$$
\end{prop}

A {\em mixed 2-quregister} is a convex combination of density matrices. 
Let ${\bf m} = \sum_{i\in I}p_i\ \rho_{2}({\bf x}_i)\in\C^{2^2\times 2^2}$ be a mixed 
state, $\forall i\in I$, $p_i\in[0,1]$ and $\sum_{i\in I}p_i=1$. 
Naturally, if, in an extreme case, for some index $i_0$ we have $p_{i_0}=1$ and $p_i=0$, for 
all $i\not=i_0$, then ${\bf m}$ is a {\em pure state}. A well known characterisation of pure states is the following:
$${\bf m}\in\C^{2^2\times 2^2} \mbox{ is pure }\ \Longleftrightarrow\ {\bf m}^2={\bf m}\ \Longleftrightarrow\ \tr({\bf m}^2)=1.$$
The mixed 2-quregister ${\bf m}$ is {\em separable} if $\forall i\in I$, ${\bf x}_i = {\bf c}_{i0}\otimes{\bf c}_{i1}$, with 
${\bf c}_{i0},{\bf c}_{i1}\in S_1(\C)$, or equivalently $\rho_2({\bf x}_i) = \rho_1({\bf c}_{i0})\otimes\rho_1({\bf c}_{i1})$.

Any separable mixed state is actually a {\em density operator}: it is symmetric, positive and with trace 1. The mixed state is determined by a $(2^2\times 2^2)$-complex matrix and such matrix determines as well a sesquilinear map $\hache_1\times\hache_1\rightarrow\C$, $({\bf x},{\bf y})\mapsto B({\bf x},{\bf y})={\bf x}^HQ{\bf y}$, which in turn determines the quadratic form ${\bf x}\mapsto Q({\bf x})={\bf x}^HQ{\bf x}$. The {\em partial traces}, regarded as quadratic forms, are
\begin{eqnarray*}
\tr_0Q:\hache_1\to\C &,& {\bf z}_0\mapsto\tr_0Q({\bf z}_0) = \sum_{k=0}^1({\bf z}_0
\otimes{\bf e}_k)^HQ({\bf z}_0\otimes{\bf e}_k), \\
\tr_1Q:\hache_1\to\C &,& {\bf z}_1\mapsto\tr_1Q({\bf z}_1) = \sum_{k=0}^1({\bf e}_k
\otimes{\bf z}_1)^HQ({\bf e}_k\otimes{\bf z}_1). 
\end{eqnarray*}
(the indexes at the argument variables ${\bf z}$, are referring to the corresponding subsystems of the composed system in $\hache_2$).
Thus, regarded as matrices, their corresponding entries are 
$$\forall i,j\in\{0,1\}: \ (\tr_0Q)_{ij} = q_{2i,2j} + q_{2i+1,2j+1}\ \ \& \ \ (\tr_1Q)_{ij} = q_{i,j} + q_{2+i,2+j}.$$

 The {\em von Neumann entropy} of the whole system is
$$E({\bf m}) = -\sum_{j=0}^3\lambda_j\log_2\lambda_j,$$
 where $\left(\lambda_j\right)_{j=0}^3$ is the collection of eigenvalues of ${\bf m}$ and, 
similarly, 
the {\em reduced von Neumann entropy} of each reduced subsystem $\tr_i{\bf m}$ is
 $$E_i({\bf m}) = -\sum_{j=0}^1\lambda_{ij}\log_2\lambda_{ij},$$
 where $\left(\lambda_{ij}\right)_{j=0}^1$ is the collection of eigenvalues of $\tr_i{\bf m}$, $i=0,1$.
The reduced von Neumann entropies entail a measure of entanglement, and, according to the Uniqueness Theorem~\cite{Don02}, sufficient and necessary conditions determine whether any other entanglement measure coincide with this criterion.
\bigskip

\noindent{\bf Example.} Consider the vector ${\bf x}_p =(\sqrt{p},0,0,\sqrt{1-p}) = \sqrt{p}\,{\bf e}_0 + \sqrt{1-p}\,{\bf e}_3\in S_3(\C)$, with $p\in[0,1]$, as in Example~\ref{sc.n1}.3. Then,
$${\bf m}_p = \rho_2({\bf x}_p) =
\left[
\begin{array}{cccc}
 p & 0 & 0 & \sqrt{1-p} \sqrt{p} \\
 0 & 0 & 0 & 0 \\
 0 & 0 & 0 & 0 \\
 \sqrt{1-p} \sqrt{p} & 0 & 0 & 1-p 
\end{array}
\right],$$
which is an idempotent matrix, ${\bf m}_p^2 = {\bf m}_p$, hence it is a 
pure state (it can be identified with ${\bf x}_p$), it has eigenvalues $0$ and $1$ of respective 
multiplicities 3 and 1, and eigenspaces ${\cal L}({\bf y}_p,{\bf e}_1,{\bf e}_2)$ 
and ${\cal L}({\bf x}_p)$ where ${\bf y}_p = (\sqrt{1-p},0,0,-\sqrt{p})$ is orthogonal to ${\bf x}_p$. The von Neumann 
entropy of ${\bf m}_p$ is thus 0. 
Now, the first partial trace is

\begin{figure}[!t]
\centering
\includegraphics[width=4in]{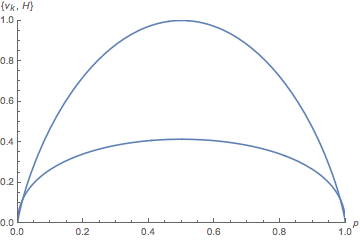} \hspace{1em} 
\caption{Graphs of $H$ and $\nu_0$. The maxima are $H(\frac{1}{2}) = 1$ and $\nu_0(\frac{1}{2}) = \sqrt{2}-1$.}
\label{fg.02}
\end{figure} 

$$(\tr_0{\bf m}_p) =
\left[
\begin{array}{cc}
 p & 0  \\
0 & 1-p 
\end{array}
\right],$$
hence $E\left(\tr_0{\bf m}_p\right) = -p\,\log_2p - (1-p)\,\log_2(1-p) = H(p)$, where $H$ is Shannon's entropy 
function. $E\left(\tr_0{\bf m}_p\right)$ is indeed a measure of the entanglement of the 2-quregister ${\bf x}_p$ 
 and it is consistent with the measure $\nu_k({\bf x}_p)$ as shown in Example~\ref{sc.n1}.3 (see Figure~\ref{fg.02}).
With this criterion, the vectors at the Bell basis correspond to maximally entangled states.
\bigskip

\section{Conclusion}

We have analysed an embedding of the unit sphere of the $2^2$-dimensional complex Hilbert space   into
the symmetry group $\mbox{\rm SU}(2^2)$. 

It is rather usual to present geometrically the collection of qubits, namely, the unit sphere $S_1(\C)$ of $\hache_1$, as the  Bloch sphere, but little attention is paid to the possible algebraic structures wthin $S_1(\C)$.

For the case $n=1$ there is a natural identification $\Psi_1$ of the sphere $S_{2}$ with $\mbox{\rm SU}(2)$, although the 
algebraic structure of $\mbox{\rm SU}(2)$ is not consistent with the application of quantum gates in $S_1(\C)$, in fact only the unitary operators that preserve orientation commute with the identification $\Psi_1$.

Unfortunately for $n\geq 2$, the natural embedding $\Psi_n$ obtained by the tensor product of the former bijection $\Psi_1$ may fail to define a bijection between $S_{2^n -1}(\C)$ and $\mbox{\rm SU}(2^n)$. In this paper, we have shown that actually for $n=2$, the embedding $\Psi_2$ does not determine a bijection between $S_3(\C)$ and $\mbox{\rm SU}(2^2)$.

However, the proposed operators  satisfy the desired embedding when they are restricted to the separable $n$-quregisters.
These operators give rise to entanglement measures which are compatible with conventional entanglement measures, as von Neumann entropy.

The procedures used in this paper are rather standard and most probably can be generalised to the quregisters of any length. We look towards to formally prove  this sketch of research.

\bibliographystyle{plain}
\bibliography{sg}
\end{document}